\documentclass[aps,prl,amsmath]{revtex4}

\usepackage{epsfig}

\renewcommand{\vec}[1]{\mathbf{#1}}  
\newcommand{\mq}{|\vec{q}|}

\newcommand{\be}{\begin{equation}}
\newcommand{\ee}{\end{equation}}
\newcommand{\bea}{\begin{eqnarray}}
\newcommand{\eea}{\end{eqnarray}}
\newcommand{\bean}{\begin{eqnarray}}
\newcommand{\eean}{\end{eqnarray*}}

\newcommand{\gapproxeq}{\lower
.7ex\hbox{$\;\stackrel{\textstyle >}{\sim}\;$}}
\newcommand{\lapproxeq}{\lower
.7ex\hbox{$\;\stackrel{\textstyle <}{\sim}\;$}}

\def\3bar{$\bar {\hbox{\bf 3}}$}
\newcommand{\qq}{$q\bar{q}$}
\newcommand{\cc}{$c\bar{c}$}

\newcommand{\epem}{e^+e^-}

\newcommand{\phalf}{\ensuremath{p_{\tfrac{1}{2}}}}

\renewcommand{\bar}[1]{\ensuremath{\overline{#1}}}

\begin{document}

\title{Possibility of Deeply Bound Hadronic Molecules from Single Pion Exchange}

\author{Frank Close}
\email[E-mail: ]{f.close1@physics.ox.ac.uk}
\affiliation{Rudolf Peierls Centre for Theoretical Physics, University of Oxford,
\\ 1 Keble Road, Oxford, OX1 3NP}
 
\author{Clark Downum}
\email[E-mail: ]{c.downum1@physics.ox.ac.uk}
\affiliation{Clarendon Laboratory, University of Oxford,\\ Parks Road, Oxford, 
OX1 3PU}

\begin{abstract}
Pion exchange in S-wave between hadrons that are themselves in a relative S-wave can shift energies by 
hundreds of MeV.
In the case of charmed mesons $D,D^*,D_0,D_1$ a spectroscopy of quasi-molecular states may arise
consistent with enigmatic charmonium states observed above 4 GeV in $e^+e^-$ annihilation.
A possible explanation of $Y(4260)\to \psi\pi\pi$ and $Y(4360) \to \psi'\pi\pi$ is found.
Searches in $D\bar{D}3\pi$ channels as well as $B$ decays are recommended to test this hypothesis.

\end{abstract}

\maketitle


The exchange of a $\pi$ in S-wave between pairs of hadrons that are themselves in relative S-wave
leads to deeply bound states between those hadrons. Examples of such spectroscopy appear to
be manifested among charmed mesons. Here we summarize the general arguments and outline a program for 
both
theoretical and experimental investigation.

 If two hadrons $A,B$ are linked by $A \to B \pi$, then necessarily hadronic pairs of $AB$ or 
$A\bar{B}$ have
the potential to feel a force from $\pi$ exchange. This force will be attractive in
at least some channels.

The most familiar example is the case of the nucleon, where $NN\pi$ coupling is the source of an 
attractive force that
forms the deuteron. Long ago\cite{torn,ericson} the idea of $\pi$ exchange between flavored mesons, in 
particular charm,
was suggested as a source of potential ``deusons"\cite{torn}. Using the deuteron binding as 
normalization, the
 attractive force between the $J^P=0^-$ charmed $D$ and its $J^P=1^-$ counterpart $D^*$ was calculated 
for the
 $D\bar D^* + c.c.$ S-wave combination with total $J^{PC}=1^{++}$, and the results compared with the 
enigmatic
 charmonium state $X(3872)$\cite{pdg08,torn,classics,fcthomas}.

Instead of normalizing the pion coupling constant to the deuteron, it may be more directly 
parametrized by decay widths, 
for example the observed width $\Gamma(D^* \to D \pi)$ \cite{fcthomas, torn}. 
Several groups have studied 
the following $J^P$ channels looking for bound states in total $J^{PC}$ channels due to pion exchange
(from here on $D\bar D^*$ etc. will be taken to include the charge conjugate channel):
\begin{eqnarray*}
D^*(1^-) \to D(0^-) \pi &\textrm{ leading to the deuson } \bar{D}D^*\; X(3872) J^{PC}=1^{++}\\
D^*(1^-) \to D^*(1^-) \pi &\textrm{ leading to the deusons } D^*\bar D^* \; J^{PC}=0^{++},1^{+-},2^{++}
\end{eqnarray*}
 These combinations were discussed in \cite{torn}. In all of these examples parity conservation 
requires that 
the $\pi$ is emitted in a P-wave; the hadrons involved at the emission vertices
 have their constituents in a relative $s$-wave, (we use S,P to denote the angular momentum between
 hadrons, and $s,p$ to denote internal angular momentum of the constituents within a hadron). 
 In such cases, the $\pi$ emission being in P-wave causes a penalty for small momentum
 transfer, ${\mathbf q}$, which is manifested by the interaction\cite{ericsonwise}
 
\begin{equation}
V_P(\vec{q}) = - \frac{g^2}{f_{\pi}^2} \frac{(\vec{\sigma}_i \cdot \vec{q})(\vec{\sigma}_j \cdot 
\vec{q})}
{\mq^2 + \mu^2}(\vec{\tau}_i \cdot \vec{\tau}_j)  .
\label{pwavepi}
\end{equation}
\noindent where $\mu^2 \equiv m_{\pi}^2 - (m_B-m_A)^2$, $m_{A,B}$ being the masses of the mesons in $A 
\to B\pi$.
(For a discussion of this interaction, and its sign, see eq 20 in ref\cite{fcthomas}).
\noindent The resulting potential is $\propto \vec{q}^2$ for low momentum transfer and has been found 
to give bindings on the scale of a few MeV, which is in part
a reflection of the P-wave penalty.

There is no such penalty when $\pi$ emission is in S-wave, which is allowed when the hadrons $A,B$
have opposite parities. Examples 
involving
the lightest charmed mesons are $D_1(1^+) \to D^*(1^-) \pi$ and $D_0(0^+) \to D(0^-) \pi$.
One might anticipate that the transition from a $D$ or $D^*$ with constituents in $s$-wave to the
$D_1$ or $D_0$
where they are in $p$-wave would restore a penalty as $\vec{q} \to 0$, leading to small binding 
effects as in the
cases previously considered. However, as we now argue, this need not be the case, and energy shifts of 
$\mathcal{O}(100)$MeV can arise.  

First, at a purely empirical level, the large widths\cite{pdg08} for $\Gamma(D_0 \to D\pi) \sim 260 
\pm 50$MeV
 and $\Gamma[D_1(2430) \to D^*\pi] \sim 385 \pm \mathcal{O}(100)$MeV imply, even after phase space is 
taken into account,
that there is a significant transition amplitude. The phenomenon of non-suppression is well known
for light hadrons and was specifically commented upon in the classic quark model paper of 
ref.\cite{fkr}. It arises from a derivative operator acting on the internal
hadron wave function, which enables an internal $s$ to $p$ transition to occur even when the momentum 
transfer
between the hadrons
vanishes. This can be seen when $\bar{\psi}\gamma_5\psi$ is expanded to $\sigma.(\vec{q}-\omega 
\vec{p}/m)$, where $\vec{q}$
 is the three-momentum transfer and $\vec{p}$ the internal quark momentum\cite{divgi}.
 Feynman, {\it et al.}\cite{fkr} argued for this on general grounds of Galilean invariance.
 The presence of $\vec{p}$ gives the required derivative operator, and hence the $p \to s$ transition.

In dynamical models of $\pi$ emission, such as the $^3P_0$ model\cite{3p0},
a \qq~must be created: $q_i \to q_i\bar{q}q \to \pi(q_i\bar{q})q$.
The creation operator is
proportional to $\sigma.\vec{p}$, where $\vec{p}$ is the momentum 
of the quark created in the \qq~ pair, which goes into the $p$-wave meson. 
In chiral perturbation theory\cite{falkluke} a similar result is found.
When applied to $\pi$ exchange in the $D^*D_1$ system (e.g. \cite{liu}) the analogue of eq. 
\ref{pwavepi}
(as presented in table 1 of ref \cite{liu}) becomes
  \begin{equation}
V_S(\vec{q}) =  \frac{h^2}{2f_{\pi}^2} \frac{(m_A-m_B)^2}
{\mq^2 + \mu^2}
\label{swavepi}
\end{equation}
The absence of a $\vec{q}^2$ penalty factor is immediately apparent, the scale now being set by
an energy gap squared, $(m_A-m_B)^2$.
  
Thus on rather general grounds we may anticipate significant energy shifts, 
$\sim \mathcal{O}(100$MeV), due to $\pi$ exchange at least in some channels 
between such hadrons in a relative S-wave.  Signals may be anticipated below 
or near threshold in the following channels (with manifest flavor, $I=1$, 
or in the charmonium analogues, $I=0$, involving charm and anticharm mesons):
\begin{eqnarray*}
  D_0(0^+) \to D(0^-) \pi &\textrm{leading to the deusons }\phantom{1}\, D\bar D_0\; 
J^{PC}=0^{-\pm}\phantom{12345}\\
  D_1(1^+) \to D^*(1^-) \pi &\textrm{leading to the deusons } D^*\bar D_1 \; J^{PC}=(0,1,2)^{-\pm}
\end{eqnarray*}
Pion exchange depends on the presence of $u,d$ flavors, therefore  
there will be no such effects in the $D_s\overline{D}_s$ analogues.
Further, the potential depends only on the quantum numbers of the light quarks.  
Therefore, there will be effects in the
strange $(K\overline{K})$ and bottom ($B\overline{B}$) analogues, which can add to the test of such 
dynamics.
  
We shall illustrate our ideas with the case of charm and draw attention to possible signals for this 
dynamics.
A variational calculation with the $I=0$~analogue of the potential of \cite{liu} suggests binding of at 
least $\mathcal{O}(100)$MeV is possible in the $I=0$ $1^{--}$ $D^*\bar D_1$ channel.  
Hence, both the ground and first radially excited eigenstates may be bound. 
As we shall show, the $Y(4260) \to \psi \pi\pi$ and $Y(4360) \to \psi'\pi\pi$ are consistent with
this picture.
We now summarize the arguments that lead to this conclusion. 

The $D_1\bar D^*$ (and charge conjugate) combination is interesting as the $1^{--}$ channel is 
accessible in $e^+e^-$ annihilation \cite{donemixing}.  
In the chiral perturbation theory model, the Fourier transformed S-wave potential has the structure 
\cite{liu}
\begin{equation}
  V_S(r) = \frac{h^2 (m_A-m_B)^2}{8 \pi f_{\pi}^2} \frac{\cos(|\mu | r)}{r}
  \label{fourier}
\end{equation}
By extracting $h$ directly from the $\Gamma(D_1 \to D^* \pi)$, following eq. (137) of 
ref.\cite{casalbuoni97}, 
we find $h = 0.8 \pm 0.1$ when $f_{\pi} = 132$MeV. 
We confirm the conclusion of \cite{liu}, that $\pi$ exchange is unlikely to form a bound state in
the $1^{--}$, $I=1$ channel.

\begin{figure}
\includegraphics[height=\textwidth,angle=-90]{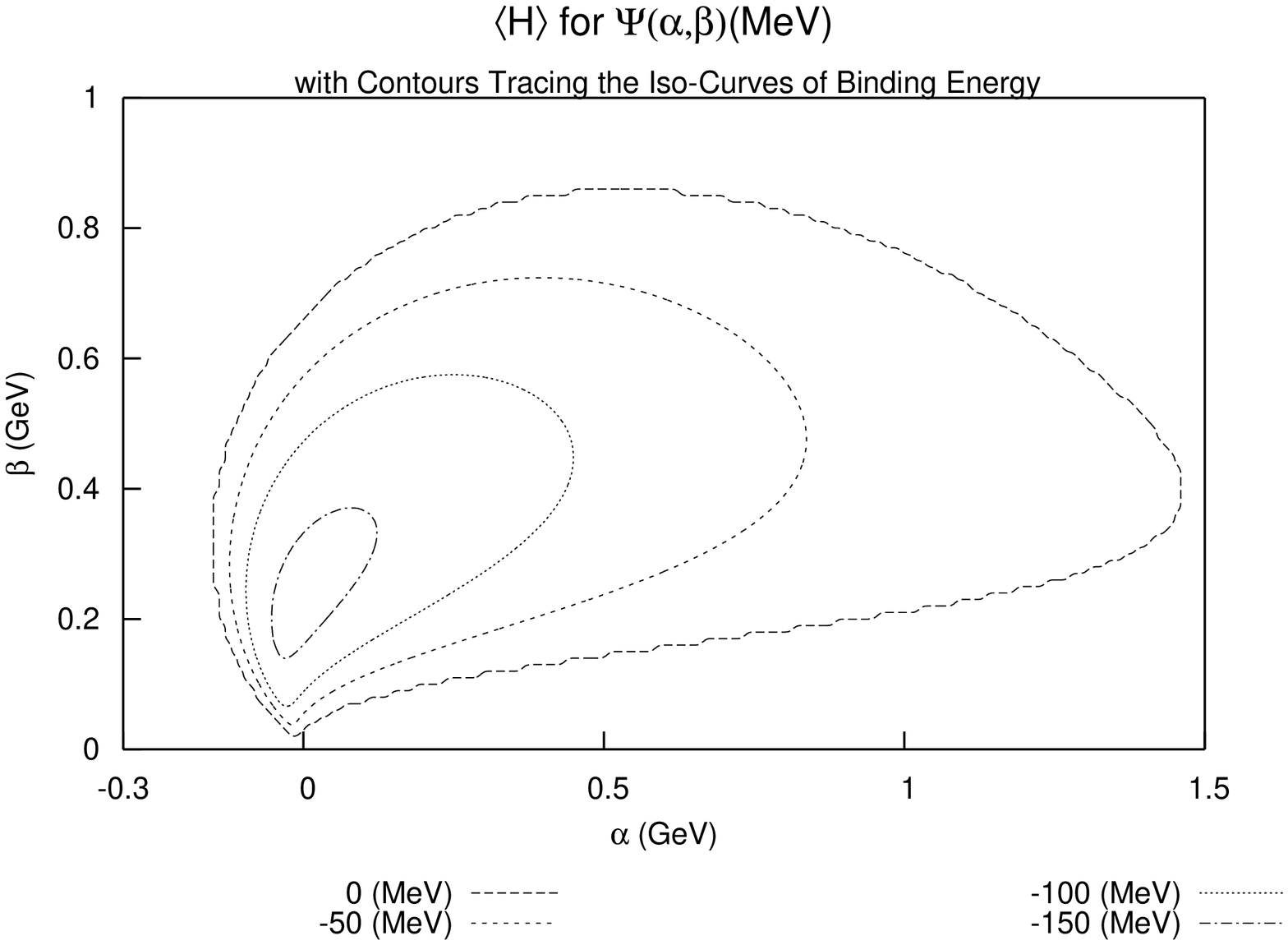}
\caption{The expectation value of the Hamiltonian with the $I=0$ potential from \cite{liu} and a trial 
wave function $\psi(r) = (1+\alpha r^2)e^{-\beta r^2}$.} 
\end{figure}
For the $I=0$ configuration things are radically changed. The $\langle {\mathbf 
\tau.\tau}\rangle $ is now of opposite sign and three 
times stronger
than for $I=1$. The opposite sign moreover causes the attractive well to occur in the leading $r \to 
0$ part of $\cos(|\mu | r)/r$.
Using a variational calculation we find a deep binding energy of $\mathcal{O}(100)$ MeV,
 (Fig 1), for the expectation value of the Hamiltonian with 
$\psi(r) = (1+\alpha r^2)e^{-\beta r^2}$, where $r$ is the
radial separation of the $D^*$ and $D_1$ mesons, and where these and the $\pi$ are 
assumed to be point like.
The variational principle
only gives lower bounds on binding energies.  A more quantitative
study which explicitly solves the Schrödinger equation will be
necessary for a precise estimate of the binding energy.
The results are robust for a wide range of physical values of $\beta$, and they are only weakly dependent 
on $\alpha$,
hence not overly sensitive to smearing, such as when modeled by form factors.
We have verified that bound states survive the ad hoc addition of a repulsive square well with a range 
of 1.25 GeV$^{-1}$ and a magnitude up to 1 GeV.
Further model dependent studies could be made but here we wish to focus on the general conclusions.

Given the depth of binding of the ground state with trial wave functions, there is the
tantalizing possibility that a radially
excited state could also be bound. As the excitation energy for radial excitation of a compact
QCD \cc~state is $\mathcal{O}(500)$MeV, it is possible that the extended molecular system may be 
excited by
less, perhaps $\mathcal{O}(100)$MeV. The rearrangement of constituents leading to
final states of the form $\psi$ + light mesons then rather naturally suggests that the lower (radial) 
state converts to $\psi \pi\pi$
( $\psi' \pi\pi$) respectively. In this context it is intriguing that there are states observed with
energies and final states that appear to be consistent with this:
$Y(4260) \to \psi \pi\pi$ \cite{4260} and the possible higher state $Y(4360) \to \psi' \pi\pi$
\cite{babar1,belle1} are respectively 170MeV and 70 MeV below the
$D^*(2010)\bar D_1(2420)$ combined masses of 4430MeV.
 
If these states were to be established as such, one could tune the model accordingly. Further, this 
could be an interesting signal for a $D^*\bar D_1$ quasi-molecular spectroscopy with transitions
among states that could be revealed in, for example, $e^+e^- \to \psi\gamma\gamma\pi\pi$.

Note also that although the $Y(4260)$ is near the $DD_{1}$ and $D^*D_0$ S-wave thresholds, parity 
conservation
forbids $\pi$ exchange other than in the off-diagonal $DD_{1} \to D^*D_0$. Thus in contrast to the
$D^*D_1$ system discussed here, $\pi$ exchange is not expected to
play a leading role in the $DD_1$ and $D^*D_0$ channels. 
As $D^*$ and $D_0 \to D\pi$ whereas $D_1 \to D \pi\pi$, the $D^*\bar D_1$ bound state 
$\to D\bar{D} 3 \pi$ in contrast to $DD_1$ or $D^*D_0 \to D\bar{D} 2\pi$.
Thus an immediate consequence 
of this interpretation is that there must be significant coupling of the $Y(4260) \to 
D\bar{D}\pi\pi\pi$
that could exceed that to $D\bar{D}\pi\pi$.

In $e^+e^-$ annihilation, the primary production is $\gamma \to c\bar{c}$ followed by hadronisation. 
This only feeds the I=0 sector. 
The $Y(4260)$ is seen also in $B \to K Y(4260)$\cite{pdg08,babarB}. This branching ratio sets the 
scale for the production of
other states in $B \to K X$. The dynamical uncertainty is whether this production is driven by a 
direct production of
two quarks and two antiquarks, or whether it is through a leading \cc~Fock state in the hadron 
wave function.
In the latter case, only $I=0$ combinations can be accessed; in effect this is like $e^+e^-$ 
annihilation but with
$b\bar{s} \to  c\bar{c}$ in any of $J^P = 0^{\pm},1^{\pm}$.
The $J^{PC} = 1^{-+}$ state can be produced via a \cc~ hybrid Fock state. The relative branching ratio 
will depend on
the relative importance of hybrid to conventional branching ratios in $B$ decays\cite{cdudek}.

Tantalizing signals for a possible resonant enhancement are seen in $B \to K\psi \omega$ immediately 
above 
threshold\cite{belle3940}.
This decay channel is allowed for the $I=0$ attractive combinations in $J^{PC} = (0,1,2)^{-+}$
We advocate that structures in $B \to K \psi 
\omega$ be investigated
and their $J^P$ determined.
   
Since the attraction of the potential depends only on the quantum numbers of the light $q\bar q$, 
it follows immediately that the flavor of the heavy quarks is irrelevant. 
Hence we expect similar effects to occur in the $b\bar{b}$ and $s\bar{s}$ sectors. 
It has been noted that $\Upsilon(10.86)$GeV
appears to have an anomalous affinity for $\Upsilon \pi\pi$\cite{upsilon}.  This state is $\sim 130$
MeV below $B^*\overline{B}_1$ threshold. In the $\phi\pi\pi$ channel there is an enhancement at 
2175MeV\cite{phi}.  
This is approximately 125MeV below the $K^*\overline K_1(1400)$ threshold.  Analysis here is more 
complex as  the heavy quark approximation fails, and the phenomenology of the $K_1(1270;1400)$
pair is more complicated\cite{barnes,lipkin}.

From theoretical and empirical studies of hadronic decay widths, 
one would expect strong S-wave pion exchange effects to occur.  These effects would be manifested in 
analogous ways in the strange, charm and bottom sectors and would provide natural explanation for 
some of the new charmonium states.  The number of bound states and the precise values of their 
energies
will be model dependent, but the deep binding
from S-wave pion exchange should be a general result.  The main quantitative uncertainty
is the experimental value for $\Gamma(D_1 \to D^* \pi)$, which sets the scale for $V_S(r)$.
Nonetheless, if $\pi$ exchange
occurs, the presence of deep binding in some channels seems unavoidable. If this is not seen, it will 
have
significant implications for the hypothesis of $\pi$ exchange as an hadronic interaction.

Further theoretical investigation is merited, in particular the roles of
form factors, sensitivity to parameters, the absorptive part of the potential due to the on-shell 
pion contribution, and to other short range forces. The physical decay of the $D_1$ into a $D^*\pi$
means that the bound state disintegrates, or even that it
fails to form.  Thus, we expect the on-shell pion contribution will
endow any state produced by this mechanism with a width, or that it
produces a non-resonant background which may obscure the signal.
However our first survey suggests that the long range
virtual $\pi$ exchange in S-wave gives a powerful attractive force, and that examples of
such a spectroscopy do appear to be manifested.

One of us (FEC) thanks Jo Dudek for a question at a Jefferson Lab seminar which stimulated some of 
this work
and T. Burns for discussion.  We would like to thank C. Thomas for comments on a draft of this work
and Dr. Yan Rui Liu for pointing out a typo.
This work is supported by grants from the Science \& Technology Facilities 
Council (UK) and in part by the EU Contract No. MRTN-CT-2006-035482, 
``FLAVIAnet.'


\begin{thebibliography}{99}
\bibitem{donemixing} The present analysis requires only that a large $D_1D^*\pi$
coupling exist for at least one of the $D_1$ states -- we take the large
experimental width from the PDG as input.  Given the near degeneracy in mass, 
the dynamical composition of this state has no effect on 
our ability to construct a $1^{--}\, D_1D^*$
molecular state and, as such, has no practical impact on this analysis.
Thus, the \phalf~notation is chosen for pedagogical reasons, not as a definitive assertion
of dynamical composition.
\end{thebibliography}


\appendix


\end{document}